\documentclass[12pt]{article}
\usepackage[latin1]{inputenc}
\usepackage{epsfig}
\usepackage{a4wide}
\usepackage{latexsym}
\usepackage{amssymb}
\usepackage{amsmath}
\begin{document}
\def\d{{\rm d}}
\def\ex{{\rm e}}
\def\u{{\bf u}}
\def\x{{\bf x}}
\def\A{{\bf A}}
\def\c{{\bf c}}
\def\W{{\bf W}}
\def\k{{\rm k}}
\def\R{{\bf R}}
\def\smalk{{\scriptscriptstyle{\rm k}}}
\def\beq{\begin{equation}}
\def\eeq{\end{equation}}
\font\brm=cmr10 at 24truept
\font\bfm=cmbx10 at 15truept
\baselineskip 0.7cm

\centerline{\brm Local evolution equations}
\vskip 5pt
\centerline{\brm  for non-Markovian processes}
\vskip 20pt
\centerline{Piero Olla$^1$ and Luca Pignagnoli$^{2,3}$}
\vskip 20pt
\centerline{$^1$ISAC-CNR, Sez. Lecce, 
% Str. Prov. Lecce-Monteroni Km 1.200,
I--73100 Lecce, Italy.}

\centerline{$^2$ISAC-CNR, I--40129 Bologna, Italy}

\centerline{$^3$Dipartimento di Matematica, Universit\'a di Milano, 
I--20133 Milano, Italy}
\vskip 20pt
\centerline{Abstract}
\vskip 5pt
A Fokker-Planck equation approach for the treatment
of non-Markovian stochastic processes is proposed. 
The approach is based on the introduction of fictitious 
trajectories sharing with the real ones their local structure
and initial conditions.
Different statistical quantities are generated
by different construction rules for the trajectories, 
which coincide only in the Markovian case. The merits and limitations
of the approach are discussed and applications to transport 
in ratchets and to anomalous diffusion are illustated.
\vfill\eject

Anomalous behaviours in the fluctuations of physical 
quantities are a common occurence, typically associated with lack of separation
between the macroscopic and the fluctuation scale \cite{bouchaud90}. 
An important example is transport in complex systems
such as disordered media \cite{sher75}, colloidal suspensions \cite{findley76}
and turbulent flows \cite{elhmaidi93}.
If a scale separation were present, the system could be 
described by a local fluctuation-dissipation process, which could be
expressed as a stochastic differential equation of the standard
Langevin type. The short-time dynamics of the fluctuating variable, call it 
$Y$, would then be characterized by normal diffusion, and there would exist
local Fokker-Planck and  backward Kolmogorov equations,
describing the dynamics of the transition PDF (probability density 
function) $\rho(Y,t|Y_0,t_0)$ \cite{schuss80}.

In general, anomalous behaviours could not be accounted for 
by a stochastic differential equation and the non-Markovian
nature of the process would be associated with equations for the PDF,
which are of integro-differential (fractional)
form in time \cite{metzler00}. This implies that, in general, 
the evaluation of statistical 
quantities will require dealing with memory kernels in the 
equation for $Y$ and with the effect of aging \cite{allegrini03}. 
However, as we shall discuss in this letter, there are
situations in which a description of the PDF dynamics,
not requiring the use of fractional equations, becomes possible.

Imagine to evaluate
the average at time $t_2$ of some function of $Y$, given the
initial condition $Y(t_0)=Y_0$. Hence, we are going to need 
the transition PDF $\rho(2|0)$ (we are going to use the 
shorthand $k$ to indicate the pair $Y_k,t_k$). 
Suppose we have already evaluated $\rho(1|0)$
for $t_0<t_1=t_2-\Delta t$ and we want to propagate $\rho(1|0)$ to $\rho(2|0)$.
Then, the following relation holds:
\beq
\rho(2|0)=\int\d Y_1\rho(2|1;0)\rho(1|0)
\label{1}
\eeq
which, in the case of a Markovian process, for which 
$\rho(2|1;0)\to\rho(2|1)$, turns into a standard Chapman-Kolmogorov
equation.

This relation could be generalized to statistical
conditions at $n\ne 1$ times. In particular,
the case $n=0$ is realized by a standard Chapman-Kolmogorov equation
in which the propagating kernel is $\rho(2|1)$. Similarly, it is 
possible to consider joint PDF's in the form 
$\rho({\bf 1}|{\bf 0})$, ${\bf 1}=\{1,1',...\}$, ${\bf 0}=\{0,0',...\}$ and
Eq. (\ref{1}) would read:
\beq
\rho({\bf 2}|{\bf 0})=\int\d Y_1\rho({\bf 2}|{\bf 1};{\bf 0})\rho({\bf 1}|{\bf 0}).
\label{1.1}
\eeq
Hence, contrary to the Markov case, generalized Chapman-Kolmogorov equations 
propagating PDF's with different conditionings at times $t_0$, $t_0'$,...
involve different transition kernels $\rho({\bf 2}|{\bf 1};{\bf 0})$.

A physical interpretation of this multiplicity is obtained observing that
an equation like (\ref{1}) 
describes the evolution of fictitious trajectories $\tilde Y(t)$ obeying
an equation in the form
\beq
\tilde Y(t_2)=\tilde Y(t_1)+\langle\Delta Y|1;0\rangle
+\Delta W
\label{2}
\eeq
with initial condition given by the pair $Y_0,Y_1$, $t_1-t_0\to 0$.
In this equation, $\Delta Y=Y(t_2|1;0)-\tilde Y(t_1)$
and $\Delta W$ has zero mean and statistics determined, in the $\Delta t\to 0$ 
limit, by $\langle\Delta W^2|1;0\rangle$. These trajectories are fictitious 
in the sense that they are not typical realizations of the process $Y$. 
In fact, in the case of a real trajectory, $\tilde Y(t_2)$ would be obtained
from $Y(t_1)$ using informations on the whole history before $t_1$,
and not only at $t=t_0$ or, in the case of Eq. (\ref{1.1}), at the discrete
instants $t_0,t_0',...$ Only in the Markov case, cease the trajectories 
to be fictitious and become typical realizations of the process,
coincident in form for different $n$. This in analogy with
Eq. (\ref{1}), which, in the Markov case, takes the unique 
standard Chapman-Kolmogorov form.

It is possible to introduce a Fokker-Planck
formalism to describe the evolution of the 1-time statistics for the fictitious
trajectories and hence of the PDF $\rho(Y,t|Y_0,t_0)$. 
To understand the form of the Fokker Planck equation associated with Eq. (\ref{1}), 
consider the case of a Gaussian stationary process with generic power law scaling at
small time separations: $C(t)=\langle Y(\tau)Y(\tau+t)\rangle
\simeq \sigma^2-B|t|^\alpha/2$, $|t|\ll(\sigma^2/B)^{1/\alpha}$. 
[This process can be turned into a fractional Brownian motion sending $\sigma\to\infty$
and allowing $C(t)=\sigma^2-B|t|^\alpha/2$ for generic $t$].
In the case of a Gaussian process, explicit expressions for 
$\langle\Delta Y|1;0\rangle$ and $\langle\Delta W^2|1;0\rangle$
can be obtained analytically \cite{olla04}:
\beq
\begin{cases}
\langle\Delta Y|1;0\rangle=\sum_{lm=0}^1C_{2l}D_{lm}Y_m-Y_1
\\
\langle\Delta W^2|1;0\rangle=C_{22}-\sum_{lm=0}^1C_{2l}D_{lm}C_{m2}
\end{cases}
\label{6}
\eeq
where $C_{ij}=\langle Y(t_i)Y(t_j)\rangle$ is the correlation matrix
at instants $t_k$, $k=0,1,2$ and $D_{ij}$ is the inverse of its restriction
to $k=0,1$: $\sum_{j=0}^1D_{ij}C_{jk}=\sum_{j=0}^1C_{ij}D_{jk}=\delta_{ik}$.

In this way we have, for $t_2=t_1+\d t$
and $t_1=t_0+t$, with $t$ generic: $C_{kk}=\sigma^2$, $k=0,1,2$,
$C_{01}=C_{10}=C(t)$,
$C_{12}=C_{21}=\sigma^2-B\d t^\alpha/2$ and
$C_{02}=C_{20}=C(t)+C'(t)\d t$.
Similarly for $D_{ij}$:
$D_{00}=D_{11}=\sigma^2\Phi(t)$ and
$D_{01}=D_{10}=-\Phi(t)C(t)$,
where $\Phi(t)=[\sigma^4-C^2(t)]^{-1}$. Substituting into Eq. (\ref{6}),
we find:
\beq
\langle\d W^2|1;0\rangle=B\d t^\alpha
\label{9}
\eeq
and
$$
\langle\d Y|1;0\rangle=\Phi(t)\{
[-\sigma^2B\d t^\alpha/2-C(t)C'(t)\d t]Y_1
$$
\beq
+[BC(t)\d t^\alpha/2+\sigma^2C'(t)\d t]Y_0\}
\label{10}
\eeq
Similar calculations could be performed also for $\langle\d W^2|1\rangle$ and 
$\langle\d Y|1\rangle$ and the result is:
\beq
\langle\d W^2|1\rangle=-\sigma^2Y_1^{-1}\langle\d Y|1\rangle=B\d t^\alpha
\label{10.1}
\eeq
We obtain from Eqs. (\ref{9},\ref{10}) the generalized Fokker-Planck equation:
\beq	
\d t_1\,\partial_{t_1}\rho+\d t_1^\beta\partial_{Y_1}(A\rho)
=\d t_1^\alpha\frac{1}{2}\partial_{Y_1}^2(B\rho),
\label{11}
\eeq
where $\beta=\min(1,\alpha)$ is the leading exponent in $\d t$ of $\langle\d Y|1;0\rangle$,
$A(1;0)=\langle\d Y/\d t_1^\beta|1;0\rangle$ plays the role of a generalized drift term and
$\rho=\rho(1|0)$.

The presence of time differentials of different order in Eq. (\ref{11})
indicates that this equation is degenerate unless $\alpha=1$, in which case also $\beta=1$.
In the subdiffusive case, we have $\beta=\alpha<1$ and the PDF dynamics is governed
by a balance between the drift and the diffusion, while the kinetic term 
$\d t_1\,\partial_{t_1}\rho$ disappears.  
In the superdiffusive case $\beta=1<\alpha$ and
the diffusion term disappears. In this case Eq. (\ref{11}) becomes a 
Liouville equation for the deterministic version of Eq. (\ref{2}): $\d Y/\d t_1=A(1;0)$ and
the probablistic content of the problem is transferred to the initial distribution for
$Y_1$ at $t_1\to t_0$.  
The Markov case 
$A(1;0)=-BY_1/2$ is recovered when $C(t)=\exp(-B|t|/2)$. The fundamental solution
$\rho(1|0)\propto\exp\{-|Y_1-Y_0|^2/(2Bt^\alpha)\}$ is obtained for $t\ll B^{-1/\alpha}$
in the three ranges $\alpha\gtrless 1$ and $\alpha=1$ \cite{note0}.

It is possible to complete the Kolmogorov pair associated with the stochastic process $Y$, 
deriving a differential equation for the conditioning variable $0$ in $\rho(2|0)$.
We consider the simpler case in which $\alpha=1$ (notice that this does not imply
that the process is Markovian). Setting $t_0=t_1-\Delta t$, $t_1<t_2$, 
the backward equation can be obtained combining the relation
$\rho(2|0)=\int\d Y_1\rho(2|1;0)\rho(1|0)$ with 
$\rho(2|1;0)=\rho(2;0|1)/\rho(0|1)$ and
$\rho(2;0|1)=\rho(0|1;2)\rho(2|1)$. The result is:
\beq
\rho(2|0)=\int\d Y_1\rho(2|1)\rho_B(1|0;2)
\label{3}
\eeq
where
\beq
\rho_B(1|0;2)=\rho(0|1;2)\frac{\rho(1|0)}{\rho(0|1)}
\label{4}
\eeq
Notice that in the Markov case, $\rho_B(1|0;2)=\rho(1|0)$ and Eq. (\ref{3}) 
takes the standard form $\rho(2|0)=\int\d Y_1\rho(2|1)\rho(1|0)$.
Invoking continuity, we can approximate the  
statistics for $\Delta Y$ as Gaussian at small time separations;
indicating $t_0=t_1-\d t$ and $t_2=t_1+t$, we can then write:
\beq
\begin{cases}
\rho(0|1;2)=c\exp\{-
\frac{1}{2}\langle\d W^2|1;2\rangle^{-1}
|\d Y-\langle\d Y|1;2\rangle|^2\}
\\
\rho(0|1)=c'\exp\{-\frac{1}{2}\langle\d W^2|1\rangle^{-1}|\d Y-\langle\d Y|1\rangle|^2\}
\\
\rho(1|0)=c''\exp\{-\frac{1}{2}\langle\d W^2|0\rangle^{-1}|\d Y+\langle\d Y|0\rangle|^2\}
\\
\end{cases}
\label{4.1}
\eeq
where $\d Y=Y_0-Y_1$. Explicit expressions for $\rho_B$ can then be obtained
substituting into Eq. (\ref{4}). In the Gaussian case described by Eqn. (\ref{9}) we 
would obtain:
$$
\rho_B(1|0;2)=c\exp\{-|\d Y-\tilde A\d t|^2/(2B\d t)\}
$$
\beq
\times
\{1+B^{-1}[2A_M\d t+\d Y^2\partial_{Y_0}]
(A_M-A)\}
\label{12}
\eeq
where $\gamma=\beta-\alpha$ and $\tilde A=A(0;2)-2A_M(0)$
with $A_M\d t=\langle d Y|0\rangle$.  Substituting 
Eq. (\ref{12}) into (\ref{3}) and Taylor expanding $\rho(2|1)$ around $Y_1=Y_0$,
a generalized backward Kolmogorov equation could then be written in explicit form
($\d t_1=-\d t$): 
\beq
[\partial_{t_1}-\tilde A\partial_{Y_1}+\frac{1}{2}B\partial_{Y_1}^2+
\hat S]\rho(2|1)=0
\label{13}
\eeq
where $\hat S=-[\partial_{Y_1}+2A_M/B](A-A_M)$. 
In the Markov case, the standard form of the backward Kolmogorov equation is
recovered: $A$ coincides with its Markovianized counterpart $A_M$, the source term $\hat S$ 
vanishes and $\tilde A=-A$. 
\vskip 10pt

\noindent{\it An application to ratchets}

\noindent A simple application of the techniques illustrated so far is the determination 
of the equilibrium statistical properties of a ratchet field, 
via the generalized Fokker-Planck equation (\ref{11}). 
Specifically, consider the uniform one-dimensional Gaussian
velocity field with correlation  $C(x,t)=\langle u(0,0)u(x,t)\rangle$:
\beq
C(x,t)=\exp\{-\frac{1}{2}[t^2+2\Lambda xt+x^2]\},
\quad\ 
|\Lambda|<1
\label{14}
\eeq
This velocity field has zero mean both in space and time, nonetheless it leads
to a non-zero mean flow; in other words, it is a ratchet \cite{reimann02}.
The drift and noise amplitude experienced by a particle
in the velocity field $u(x,t)$ is obtained following the same procedure
leading from Eqs. (\ref{6}) to (\ref{9},\ref{10}), just neglecting 
the initial condition at $t_0$ and defining 
$C_{ij}=\langle u(x(t_i),t_i)u(x(t_j),t_j)\rangle$. The result has a form 
analogous to Eq. (\ref{10.1}):
\beq
\langle\d W^2|u\rangle =
-2u^{-1}\langle\d u|u\rangle
=[1+2\Lambda u+u^2]\d t^2
\label{15}
\eeq
thus, the stochastic process $u(t)$ is ballistic at short time scales.
The equilibrium PDF of a particle travelling with
velocity $u(x(t),t)$ will obey the generalized Fokker-Planck equation
\beq
\partial_u(\langle\d u|u\rangle\rho)=\frac{1}{2}\partial_u^2(\langle\d W^2|u\rangle\rho)
\label{16}
\eeq
whose solution, from Eq. (\ref{15}) is:
$$
\rho(u)=c[1+2\Lambda u+u^2]^{-1}\exp(-u^2/2)
$$
The ratchet mean flow will be given by the first moment of this distribution, plotted in
Fig. 1.
%%%%%%%%%%%%%%%%%%%%%%%%%%%%%%%%%%%%%%%%%%%%%%%%%%%%%%%%%%%%%%%%%%%%%
\begin{figure}
\begin{center}
\includegraphics[draft=false, scale=0.9, clip=true]{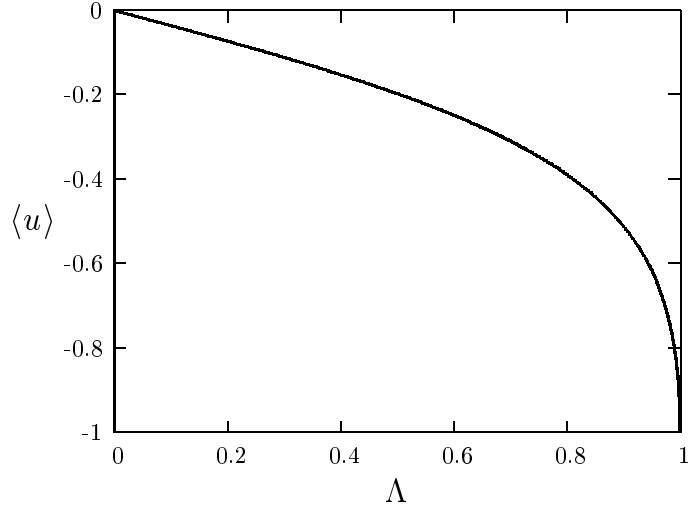} 
\caption{Flow velocity of the ratchet field of Eq. (\ref{14})}
\label{montefig1} 
\end{center}
\end{figure}
%%%%%%%%%%%%%%%%%%%%%%%%%%%%%%%%%%%%%%%%%%%%%%%%%%%%%%%%%%%%%%%%%%%%%

\vskip 10pt
\noindent{\it An application to anomalous diffusion}

\noindent As a second application, we determine the correlation between velocity and coordinate
in a time and space homogeneous, unbiased, but otherwise generic diffusion process.
In this case, $Y(t_1)$ indicates the coordinate of the walker with $Y(t_0)\equiv Y(0)=0$, and
$\rho(1|0)$ is the distribution of the walkers at time $t_1$.
We have also:
\beq
\langle|Y(t_2)-Y(t_1)|^2\rangle=\langle B(1;0)|1\rangle\Delta t^\alpha=B(1)\Delta t^\alpha
\label{17}
\eeq
with $B(1)=B$  constant from homogeneity of the process.  The velocity
averaged at scale $\Delta t$ is $\Delta Y/\Delta t$; using the definition 
$A(1;0)=\langle\d Y/\d t^\beta_1|1;0\rangle$, with $\beta=\min(1,\alpha)$ [see Eq. (\ref{11})], 
we have:
\beq
\langle Y\Delta Y/\Delta t\rangle=\Delta t^{(\beta-1)}\langle YA\rangle
\label{17.1}
\eeq
We have seen that, in the three regimes $\alpha\gtrless 1$ 
and $\alpha=1$, the generalized
Fokker-Planck equation (\ref{11}) takes the form:
$$
\begin{cases}
\partial_{t_1}\rho(1|0)+\partial_{Y_1}[A(1;0)\rho(1|0)]=0,\quad \alpha>1
\\
\partial_{t_1}\rho(1|0)+\partial_{Y_1}[A(1;0)\rho(1|0)]
=\frac{1}{2}\partial_{Y_1}^2[B(1;0)\rho(1|0)],\quad \alpha=1
\\
A(1;0)\rho(1|0)=\frac{1}{2}\partial_{Y_1}[B(1;0)\rho(1|0)],\quad \alpha<1
\end{cases}
$$
Multiplying these equations by appropriate powers of $Y(t_1)$ and taking averages, 
we obtain, using Eqs. (\ref{17},\ref{17.1}):
\beq
2\langle Y\Delta Y/\Delta t\rangle=
\begin{cases}
\alpha B t^{\alpha-1},& \alpha>1
\\
0,& \alpha=1
\\
-B\Delta t^{\alpha-1},&\alpha<1
\end{cases}
\label{18}
\eeq
which gives a quantitative content to the concepts of persistence and antipersistence
in generic diffusion processes.

\vskip 10pt
Further applications of the present Fokker-Planck approach are limited
by the fact
that several operations, natural for Markovian processes, become tricky 
in the general case. The central issue appears to be the multiplicity of 
drift and diffusion coefficients generated by different conditioning choices. 

This has the important consequence that the only physical solutions of 
equations like (\ref{11}) are those at statistical equilibrium.  For 
instance, if we considered the version of Eq. (\ref{11}) obtained from the 
unconstrained moments of Eq. (\ref{10.1}), we would obtain an evolution 
equation for the one-time PDF \cite{note}, which could be solved for an arbitrary, out 
of equilibrium initial condition $\tilde\rho_0(Y)$.
This, however, would produce a Markovianized time dependent
statistics loosing all the scaling properties of the original process.
The right evolution is given by: 
$$
\tilde\rho(1)=\int\d Y_0\rho(1|0)\tilde\rho_0(Y_0)
$$
In other words, conditioning is necessary to determine the 
evolution of an out of equilibrium one-time PDF.  On the
same line of reasoning, we find that the evolution described by
Eqs. (\ref{9},\ref{10},\ref{11}) (that is conditioned only 
at one time) is unable to account for the 
aging of the process, which may be defined, including the possibility of
non-renewing
processes, as the approach to equilibrium of the correlation
$\langle Y(t_2)Y(t_1)|0\rangle$ as $t_0\to -\infty$. In
this case, taking $t_2>t_1$,
\beq
\langle Y(t_2)Y(t_1)|0\rangle=
\langle\langle Y(t_2)|1;0\rangle Y(t_1)|0\rangle
\label{18.5}
\eeq
would require evaluation of $\langle Y(t_2)|1;0\rangle$ by means of
a version of the generalized Fokker-Planck equation (\ref{11}) conditioned 
at two times.

{\it Thus, in general, to determine non-equilibrium statistics conditioned 
at $n$ times, it is necessary to consider a generalized Fokker-Planck
equation with moments conditioned at $n+1$ times.}

These problems limit also the applicability of a Monte Carlo
approach based on the fictitious trajectories defined in Eq. (\ref{2}).
Consider for instance the motion of a particle moving with velocity
$Y$. One may try to obtain the displacement statistics by Monte Carlo
integration of the generalized Langevin equation (\ref{2}), coupled with
the kinematic condition on the particle coordinate $x(t)$: $\d x=Y\d t$. 
The first moment of the displacement is evaluated correctly:
$$
\langle x(t)|0\rangle=x(0)+
\int_0^t\d\tau_1\langle Y(t_1)|0\rangle
$$
and $\langle Y(t_1)|0\rangle$ is generated averaging over the trajectories determined
by Eq. (\ref{2}).
But, suppose we wish to calculate the second moment; we have:
$$
\langle [x(t)-x(0)]^2\rangle
=2\int_0^t\d\tau_1\int_{\tau_1}^2\d\tau_2\langle Y(t_1)Y(t_2)\rangle
$$
with
\beq
\langle Y(t_2)Y(t_1)\rangle=\langle\langle Y(t_2)|1;0\rangle Y(t_1)\rangle.
\label{19}
\eeq
The situation is analogous to Eq. (\ref{18.5}) and the conditional average
$\langle Y(t_2)|1;0\rangle$ is evaluated correctly by a Monte Carlo with
conditioning at a single time, only in the 
Markov case, when $\langle Y(t_2)|1;0\rangle=\langle Y(t_2)|1\rangle$. 

What happens is that $x(t)$ depends on the whole history of $Y$.
To obtain the full $x(t)$ statistics, the conditional moments entering Eq. 
(\ref{2}) should be substituted at time $\tau<t$ by others depending not only 
on $Y(\tau)$ and $Y(0)$, but also on $x(\tau)$, which is equivalent to 
adopting a non-local approach like the one in \cite{metzler00}.  
(Hence, the result in \cite{olla04} on the Lagrangian correlation time
in a uniform Gaussian velocity field, which neglects this conditioning, has at 
most value of estimate).

\vskip 10pt
Another set of questions relate to the existence of a stochastic process, for a given
choice of drift and diffusion coefficients, and, conversely, to the uniqueness
of the generalized Fokker-Planck equation associated to a given process.

As regards the first question, at least if we want to construct a stationary 
process, it turns out that the choice of drift and diffusion coefficient in 
Eq. (\ref{10.1}) is not free, and the functions $A(1;0)$, $B(1;0)$ have to
be chosen together with the one- and two-time PDF's $\rho(1)$ and $\rho(1|0)$:
it is not enough to impose that $A$ and $B$ depend solely on time differences.

In fact, from $A(1;0)$ and $B(1;0)$, if the process were stationary, 
it would be possible, taking the limit $t_1-t_0\to \infty$ 
in $A(1;0)$ and $B(1;0)$, 
to calculate the unconditioned moments $A(1)$ and $B(1)$,
and, using the unconditioned version of Eq. (\ref{11}), the one-time equilibrium
PDF $\rho(1)$. At the same time, knowing $A(1)$ and $B(1)$ would give the form
of the transition PDF $\rho(1|0)$ for $t_1-t_0\to 0$, to be used as
initial condition for Eq. (\ref{11}) in the calculation of $\rho(1|0)$ at finite
time separations. Under these conditions, $\rho(1)$ and $\rho(1|0)$ would be 
determined by the pair $\{A(1;0),B(1;0)\}$ in the two regimes in which 
$t_1-t_0$ is infinite and finite, respectively. It would then be easy to 
construct profiles of $A(1;0)$ and $B(1;0)$, fixed $A(1)$ and $B(1)$, 
such that the condition
$\rho(1)=\int\rho(1|0)\rho(0)\d Y_0$ be not satisfied for finite
$t_1-t_0$, which is absurd.

In conclusion, contrary to the Markov case, to be able to write down 
a generalized Fokker-Planck equation, it would be necessary to know in
advance the form of the transition PDF that is its solution; a 
characteristic shared by the approach in \cite{adelman76}.
The main content of an equation like (\ref{11}) seems therefore just to 
establish a relation, which should 
be satisfied by any experimentally measured stochastic time series, between the 
transition PDF $\rho(1|0)$ and the conditional moments $\langle\d Y|1;0\rangle$ 
and $\langle\d Y^2|1;0\rangle$. 

Turning to the uniqueness issue, we observe that the generalized Fokker-Planck equation
(\ref{11}), given a transition PDF $\rho(1|0)$, does not fix by itself the 
form of the drift and diffusion coefficients $A(1;0)$ and $B(1;0)$. 
In order for that equation to have a probabilistic content, it is necessary that 
the consistency condition 
\beq
\langle g|1\rangle=\int\d Y_0\langle g|1;0\rangle\rho(0|1)=
\int\d Y_0\langle g|1;0\rangle\frac{\rho(1|0)\rho(0)}{\rho(1)}
\label{20}
\eeq
be satisfied for $g=\Delta Y,\Delta W^2$. However, this condition, together with Eq. (\ref{11})
is still not sufficient to fix $A(1;0)$ and $B(1;0)$.

To prove this, the following  simple example is sufficient. Consider a space and 
time homogeneous, unbiased diffusion 
process. Again, $Y$ indicates the walker coordinate. From
space homogeneity, $\langle\Delta W^2|1;0\rangle$ and $\rho(0|1)$ must be
function only of $Y(t_1)-Y(t_0)$ and, from the process being unbiased,
both functions must be even. Again from the process being unbiased, $\langle\Delta Y|1;0\rangle$
must be an odd function of $Y(t_1)-Y(t_0)$ and $\langle\Delta Y|1\rangle$ must be
identically zero. We see then that the definition
$$
\int\d Y_0\langle\Delta W^2|1;0\rangle\rho(0|1)=
\langle\Delta W^2|1\rangle=B\Delta t^\alpha
$$
is satisfied by the general solution
$$
\langle\Delta W^2|1;0\rangle=B\{1+c[1-\frac{g(1;0)}{\langle g|1\rangle}]\}\Delta t^\alpha
$$
where $g(1;0)=g(Y_1-Y_0,t_1-t_0)$ is generic and
$c$ is chosen so that $\langle\Delta W^2|1;0\rangle$ remains positive defined.
Substituting into the generalized Fokker-Planck equation  (\ref{11}), and solving for
$A(1;0)$, we see that the result is an odd function of $Y(t_1)-Y(t_0)$, so that
$A(1)=\langle A(1;0)|1\rangle$ remains zero independently of $g$.

Thus, uniqueness is not guaranteed in general. However, in order for Eq. (\ref{20}) 
to be satisfied, it is still  necessary
that the moments of $\rho(2|1;0)$ and $\rho(2|1)$ be of the same order in $\d t$. 
This property is not satisfied by the Fokker-Planck equation derived in
\cite{adelman76}, assuming 
that, irrespective of $\alpha$, $\langle\d Y|1;0\rangle$ and $\langle\d W^2|1;0\rangle$
be both $O(\d t)$. Like the approach in this letter, also the one in
\cite{adelman76} can be seen as a reconstruction
technique of the transition PDF $\rho(1|0)$, based on the use of fictitious trajectories.
In the case of \cite{adelman76}, 
however, not only have the trajectories a memory of the past limited to 
the discrete time $t_0$, but, imposing the
local Poisson-like behavior $\d Y\sim \d t^{1/2}$, they
loose also all information on the local structure of the real trajectories. 
This clearly precludes any application of the type leading to Eq. (\ref{18}).

\vskip 5pt
This work was supported by the Commission of the
European Communities under contract EVK2-2000-00544 of the
Environment and Climate Programme.


\begin{thebibliography}{99}
\bibitem{note0} 
It is clear that identical scaling behaviors 
at finite time separations could be obtained irrespective of the value of $\alpha$.
Conversely, the differential $\d t$ is taken above the microscopic scale identifying
the end of the scaling range.
\bibitem{note} 
This is true, provided the kinetic piece $\partial_t\rho$ survives the elimination
of the terms which are of higher order in $\d t$.
\end{thebibliography}
\end{document}